\documentclass[12pt]{iopart}
\usepackage{graphicx}
\begin{document}

\title{
Noise and Thermal Depinning of Wigner Crystals 
} 
\author{
C. Reichhardt and C. J. O. Reichhardt 
} 
\address{
Theoretical Division and Center for Nonlinear Studies,
Los Alamos National Laboratory, Los Alamos, New Mexico 87545, USA
} 
\vspace{10pt}
\begin{indented}
  \item[]{\today}
\end{indented}

\begin{abstract}
We examine changes in the depinning threshold and conduction noise fluctuations for driven Wigner crystals in the presence of quenched disorder. At low temperatures there is a well defined depinning threshold and a strong peak in the noise power with $1/f$ noise characteristics. At higher temperatures, the depinning threshold shifts to lower drives and the noise, which is also reduced in power, becomes more white in character. At lower temperatures, a washboard frequency appears when the system depins elastically or forms a moving smectic state; however, this washboard signal is strongly reduced for higher temperatures and completely disappears above the melting temperature of a system without quenched disorder. Our results are in good agreement with recent transport and noise studies for systems where electron crystal depinning is believed to arise, and also show how noise can be used to distinguish between crystal, glass, and liquid phases.  
\end{abstract}

\section{Introduction}

When a collection of interacting particles 
is driven over quenched disorder, the system 
can exhibit a pinned phase, a depinning threshold,
and a sliding phase \cite{Fisher98,Reichhardt17}. The
existence of these phases
can be deduced from changes in transport measures
such as the velocity-force and 
differential resistance curves
\cite{Fisher98,Reichhardt17,Gruner81,Gruner88,Pertsinidis08,Bohlein12,Schulz12,Vanossi13}.
If the particles maintain the same neighbors during
the depinning and sliding process, the depinning is considered elastic
and is associated with specific scaling features 
in the velocity-force curves
\cite{Fisher98,Reichhardt17,Reichhardt02,DiScala12}, while if there is
tearing or mixing of the particles, the behavior is 
plastic and can produce multiple steps or jumps in the transport curves 
\cite{Fisher98,Reichhardt17,Reichhardt02,Fily10}.
In the sliding phase, there can also be dynamical transitions 
between different types of plastic flow or fluidlike flow,
as well as dynamical ordering transitions where the
driven particles move so rapidly over the substrate
that the effectiveness of the pinning is reduced 
and a disordered system can organize into 
a moving crystal or smectic
state
\cite{Reichhardt17,Bhattacharya93,Koshelev94,Moon96,Giamarchi96,Balents98,Olson98a,Pardo98,Kolton99,Reichhardt15}. 
When thermal effects are included, additional behaviors can occur
both during depinning and in the sliding states. In general,
sharp depinning thresholds become rounded due to thermal creep; 
however, a peak in the differential velocity can still arise
near the $T = 0$ depinning threshold
due to a transition from creep to sliding dynamics
\cite{Fisher98,Reichhardt17,Koshelev94}.
If the temperature is higher than the melting temperature of the
quenched disorder-free system,
a system containing quenched disorder will always be in a
disordered state,
and can form a glass phase with thermal creep or
a fluctuating liquid state at high drives
\cite{Reichhardt17,Koshelev94,Hellerqvist96,Ryu96}. 

Another method to examine the driven dynamics 
is to measure changes in the noise for systems in which
time series of the velocity or density fluctuations
can be obtained as a function of drive 
\cite{Reichhardt17,Gruner88,Olson98a,Weissman88,Bloom93,Marley95,Okuma07,Bullard08,Okuma08,Liu18,Diaz17,Bennaceur18,Sato19,Sun22}.
One of the most common characterization techniques is
to determine the power spectrum of the fluctuations
and to measure the noise power over some frequency range.
For elastic depinning or ordered moving states
where the particles maintain a fixed set of neighbors and
travel at speed $v$, there is 
typically a narrow band noise signature
containing peaks at specific frequencies $\omega=2\pi v/a$
that are associated with the
average spacing $a$ between the particles
\cite{Reichhardt17,Gruner88}.
Additional peaks appear at higher harmonics of these characteristic
frequencies.
Narrow band noise signatures have been observed for sliding charge 
density waves \cite{Gruner88},
superconducting vortex lattices
\cite{Olson98a,Okuma07,Bullard08,Okuma08,Maegochi22}, 
moving charge crystals \cite{Sun22}, and skyrmion systems
\cite{Diaz17,Sato20}. 
In some cases it is possible to observe
multiple frequencies when the system is broken up into 
several large sections that locally behave elastically but
globally provide multiple degrees of freedom, permitting switching
behavior to occur
\cite{Sun22,Csathy07}.
If the depinning is strongly plastic, the 
narrow band noise signal is lost and is typically replaced by
$1/f^{\alpha}$ noise with $0.75 < \alpha < 2.0$
\cite{Reichhardt17,Olson98a,Marley95,Bullard08,Okuma08},
while for a fluid the noise power is often low and the fluctuations have
white noise characteristics with $\alpha = 0$.
In other cases, the fluctuations are Lorentizan and the
noise has a $1/f$ shape at low frequencies but becomes white
above a characteristic frequency $\omega_c$ that is
associated with the average time between collisions of
particles with pinning sites
\cite{Reichhardt17}. 
Broad band $1/f$ noise has been observed
near the depinning transition for superconducting vortices 
\cite{Reichhardt17,Olson98a,Marley95,Okuma08},
magnetic skyrmions \cite{Diaz17,Sato19},
and charge crystals \cite{Sun22,Cooper03}.  
The noise measurements
can also be used to identify a transition between different
dynamical states such as plastic flow to dynamically ordered flow.
In this case, $1/f$ noise occurs in the plastic
flow phase just above depinning, 
but at higher drives there is a
crossover to narrow band noise
as dynamical ordering emerges \cite{Reichhardt17,Olson98a,Okuma08,Diaz17}.
These different noise features can be modified significantly
when thermal effects become important \cite{Reichhardt17}. 

Another example of an assembly of particle-like objects
that can be coupled to quenched disorder and driven 
is electron crystals or Wigner crystals
\cite{Wigner34,Monceau12,Shayegan22,Andrei88,DaSilva10},
where transport measures provide evidence for a conduction threshold 
that is consistent with the existence of a
depinning transition
\cite{Monceau12,Shayegan22,Goldman90,Williams91,Jiang91,Li91,Csathy07,Kopelevich07,Hossain22}.
Recently, a growing number of materials have been
identified that can support Wigner crystals,
such as moir{\` e} superlattices \cite{Xu20,Li21},
transition metal dichalcogenide monolayers
\cite{Smolenski21,Zhou21,Matty22}, and 
systems where Wigner crystals are stable
at zero magnetic field \cite{Falson22}.
It would be interesting to examine conduction and noise measures
as a function of drive and temperature in these new systems. 
Previous experiments that showed evidence of a
conduction threshold also revealed a large increase
in the conduction noise just above depinning \cite{Li91}, 
and previous numerical studies of driven Wigner crystals
also showed both a conduction threshold and
$1/f$ noise features
near depinning followed by a crossover to narrow band
noise at higher drives \cite{Reichhardt01}. 
Brussarski {\it et al.}
\cite{Brussarski18} examined the transport and
noise of Wigner crystals near depinning 
as function of temperature, and
found that at low temperature, there is a sharp depinning threshold 
that is correlated with a large peak in the noise power. 
Additionally, the noise near depinning
is of $1/f^{0.75}$ form.
As the temperature is increased, the
depinning threshold shifts to lower values
and the peak noise power is also reduced. 
This suggests that at higher temperature, the system forms
a Wigner liquid in which
the correlated motion associated with glassy or plastic flow 
phases and large noise power is lost.
Noise studies have also been performed near the metal-insulator
transition, which could be associated with a change from a Wigner glass to
a Wigner liquid,
and a drop in the noise power is observed
at higher temperatures where a fluid phase may be present
\cite{Bogdanovich02,Jaroszynski02}.
Particle-based simulations across a Wigner glass to Wigner fluid crossover
show high power $1/f^\alpha$ noise in the Wigner glass state
and lower noise power with a white spectrum at higher temperatures
in the fluid state \cite{Reichhardt04}.
Thermal effects and thermal melting in
Wigner crystals have also been extensively studied
\cite{Chen06,Knighton18,Deng19,Ma20,Kim22},
so it should be feasible to perform experimental
noise and transport measures
across a thermal melting transition while the system is being 
driven.

In this work, we consider thermally induced transport and noise measurements
for a two-dimensional (2D) electron system driven over quenched disorder.
Previous work on this system
focused on the $T = 0$ case, and showed that for plastic depinning, there
is strong $1/f$ noise with a peak in the noise power near
the depinning transition, followed by a
drop in the noise power and a transition to white or narrow band noise
at high driving where a moving smectic or moving crystal phase
emerges \cite{Reichhardt01,Reichhardt22}. 
Here we find that as we increase the temperature, 
the depinning threshold decreases
and the noise power drops,
in agreement with experiments.
Additionally, we find that the 
narrow band noise visible for $T = 0$ at high drives
is strongly 
reduced at higher temperatures and vanishes
above the temperature $T_m$ at which the system melts in the absence of
quenched disorder.
This suggests that narrow band noise signals may only be accessible
at temperatures well below melting.
We map out the dynamic phase diagram
as a function of drive versus temperature
and show that at $T_m$ there is a divergence
in the drive at which a transition to ordered or partially ordered flow occurs,
similar to the dynamic phase diagram
proposed
by Koshelev and Vinokur for driven
superconducting vortex systems \cite{Koshelev94,Ryu96}. 
For the case of elastic depinning,
we find a thermally induced creep regime 
in which the lattice moves by one lattice constant at a time,
and show
that a narrow band signal can still arise even in the creep 
regime. The spectral peaks become sharper and shift to 
higher frequencies with increasing drive,
but the narrow band signature is lost with increasing temperature
even before the system reaches the clean melting temperature $T_m$.

\section{Simulation and System}

We model a 2D classical Wigner crystal 
with charge density 
$n = N_{e}/L^2$, where $N_{e}$ is the number of electrons and $L$ is the system
size. We employ periodic
boundary conditions in the $x$ and $y$ directions, and
the sample contains $N_p$ randomly placed
pinning sites
modeled as short range attractive wells
with a density of $n_{p} = N_{p}/L^2$.
Throughout this work we fix $n=0.208$ and $n_p=0.25$.
At $T = 0$ and in the absence of quenched disorder,
the charges form a triangular lattice that has a 
well defined melting transition temperature $T_{m}$ \cite{Reichhardt22a}. 
Additionally, when $T = 0$ there is a well defined quenched disorder 
strength above which the system disorders \cite{Reichhardt22a}.
We represent the charges using
a previously studied model
\cite{Reichhardt01,Qian17,Reichhardt22,Reichhardt22a,Cha94a,Cha95,Piacente05,Reichhardt21}, where  
the equation of motion for charge $i$ is 
\begin{equation} 
\alpha_d {\bf v}_{i} = \sum^{N}_{j}\nabla U(r_{ij}) +  {\bf F}_{p} + {\bf F}_{D} 
+ {\bf F}^{T}_{i}  \ .
\end{equation}
Here $\alpha_{d}$ is a damping term  and
$U_{i} = q^2/r$ is the long range Coulomb repulsion between charges
of magnitude $q$.
As in previous work
\cite{Reichhardt01,Reichhardt21}, we employ 
a Lekner summation
to evaluate the long range interactions.
The second term on the right hand side represents pinning sites
modeled as finite range parabolic traps that
impart a maximum pinning force of $F_{p}$ at radius $r_{p}$. 
The thermal fluctuations are applied with
the term ${\bf F}^{T}$, which 
has the following properties:
$\langle F^{T}\rangle = 0$ and
$\langle F^{T}(t_i)F^{T}(t_j^{\prime})\rangle = 2k_BT\delta_{ij}\delta(t-t^\prime)$.  
The initial positions of the charges are obtained through
simulated annealing at zero drive.
Once the system has been initialized,
we apply a driving force ${\bf F_D}=F_D{\bf \hat x}$ representing
an applied voltage.
The drive can be set to a constant value, in which case
we wait for the system to reach a steady state before
measuring the average velocity per charge
$\langle V\rangle = \sum^{N_e}_i{\bf v}_i\cdot {\hat {\bf x}}$ 
or obtaining a time series of the velocity 
to examine the temporal fluctuations. 
By considering a range of drives and measuring
the average velocity at each drive,
we can create a current-voltage curve. 
If there is a magnetic field present,
the changes experience an additional 
force $q{\bf B}\times {\bf v}_{i}$ that can create a
Hall angle for the electron motion. This effect is generally  
small and we neglect it in the present work, but we have
studied it in detail
elsewhere \cite{Reichhardt21}.

\section{Elastic and Plastic Regimes}

\begin{figure}
\includegraphics[width=\columnwidth]{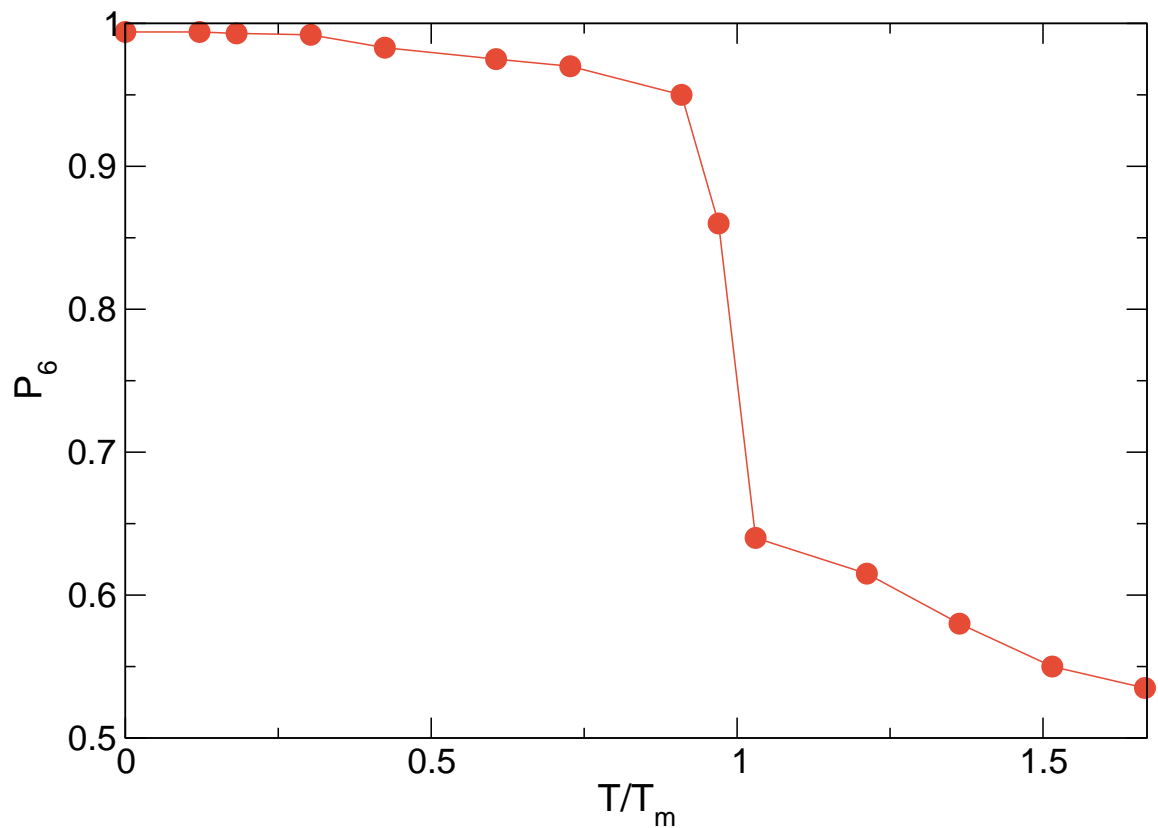}
\caption{
The average fraction of sixfold-coordinated charges $P_6$
versus temperature $T/T_{m}$ in a system with
no quenched disorder.
$T_{m}$ is defined to be the temperature at
which a proliferation of non-sixfold coordinated charges
occurs in a clean system.
}	
\label{fig:1} 
\end{figure}

In Fig.~\ref{fig:1} we plot the fraction $P_6$
of six-fold coordinated charges
versus temperature $T/T_m$ for
a system with
no quenched disorder.
The melting temperature $T_{m}$ is defined to be the temperature at
which a proliferation
of topological defects or non-sixfold coordinated charges occurs.
For $T/T_{m} < 1.0$, $P_{6}$ is close to $1.0$, as expected
for a triangular lattice, while for $T/T_{m} > 1.0$, 
a large number of fivefold and sevenfold coordinated charges appear,
causing $P_6$ to drop.

\begin{figure}
\includegraphics[width=\columnwidth]{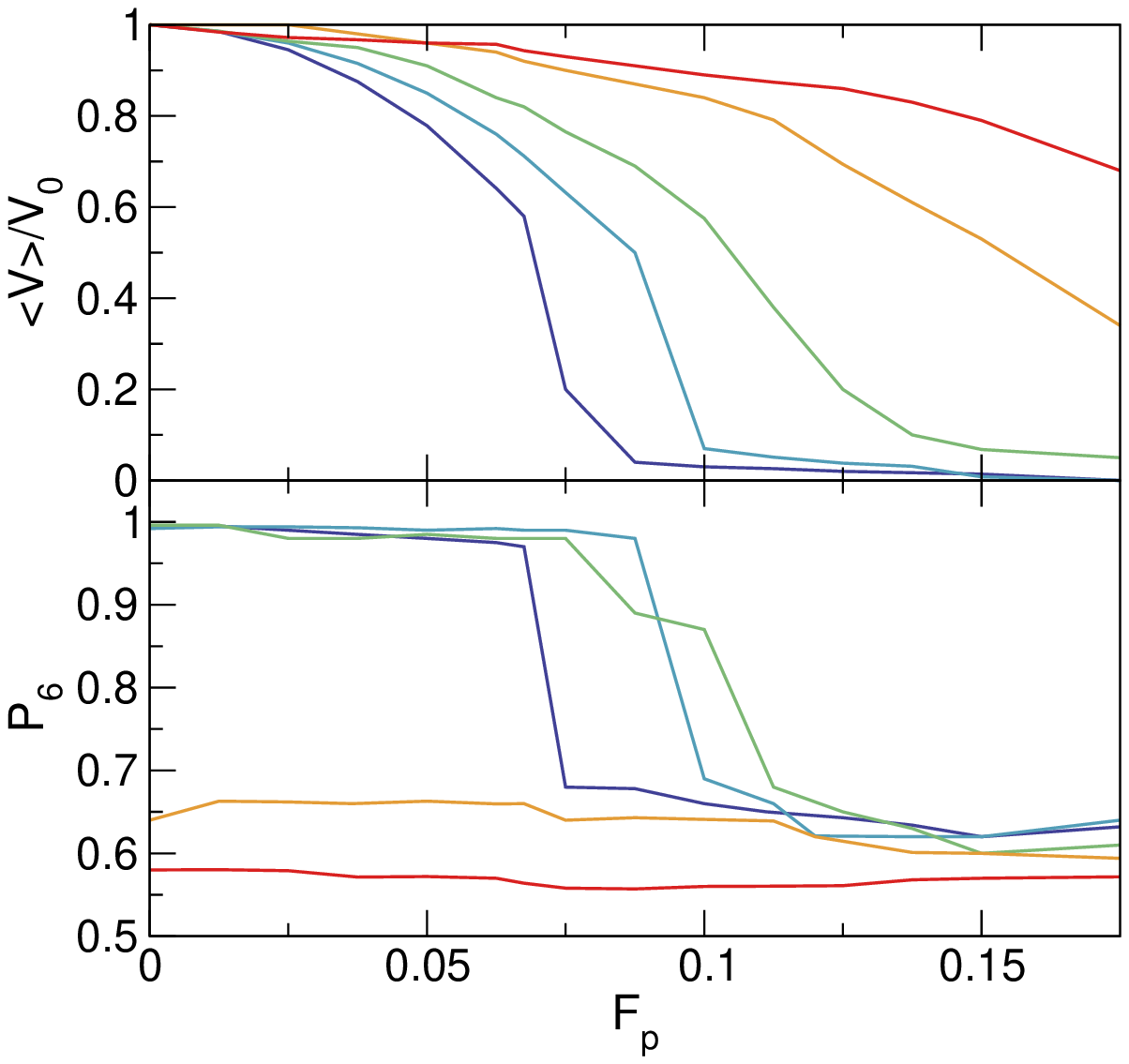}
\caption{(a) The average charge velocity $\langle V\rangle/V_{0}$ vs
pinning strength $F_{p}$ at $F_D=0.01$,
where
$V_0=F_D=0.01$ is the average velocity in a disorder-free system.
$T/T_{m} = 0$ (dark blue), $0.3$ (light blue), $0.6$ (green),
$1.03$ (yellow) and $1.38$ (red).
(b) The corresponding $P_{6}$ vs $F_{p}$
showing that for $T/T_{m} < 1.0$,
there is a well defined pinning-induced order to disorder transition.}
\label{fig:2}
\end{figure}

Once we have defined $T_{m}$ by measuring a clean system, we
introduce quenched disorder
in order to
study the conduction noise and transport response
above and below $T_{m}$ for varied disorder strengths $F_p$.
We apply a constant drive with $F_{D} = 0.01$
to samples with different $F_p$ and 
measure the time average velocity per charge
$\langle V\rangle$ 
over $4\times 10^6$ simulation
time steps.
When $F_{p} = 0$, the charge velocity $V_0$ is identical to the
driving force, $V_0=F_D=0.01$, so a measurement of
$\langle V\rangle/V_0=1$ indicates that the flow of the charges
has reached the pin-free limit.
In Fig.~\ref{fig:2}(a), where we plot $\langle V/V_0\rangle$
versus $F_p$, at $T/T_m=0$
there is a large drop in
$\langle V\rangle/V_{0}$ near $F_{p} = 0.75$. 
Figure~\ref{fig:2}(b) shows the corresponding values
of $P_{6}$ versus $F_{p}$, where for 
$T/T_{m} = 0$ there is a well defined transition from an ordered Wigner 
crystal to a disordered Wigner glass, and 
the proliferation of defects correlates with the 
velocity drop in Fig.~\ref{fig:2}(a).
At $T/T_{m} = 0.3$, the overall velocity
is higher than
for the $T/T_{m} = 0$ sample due to the lowering of the effectiveness
of the pinning by the thermal fluctuations.
Additionally,
the pinning strength required to disorder the 
system is shifted upward to a value
close to $F_{p} = 0.1$, which is again due to the partial reduction
of the pinning effectiveness by the
thermal fluctuations.
A similar trend occurs for $T/T_{m} = 0.6$, where the
velocity is higher.  For $T/T_{m} = 1.03$,
the system is disordered for all values of $F_{p}$
and the velocity is even higher but has a
gradual drop with increasing $F_{p}$,
and the same trend occurs for
$T/T_{m} = 1.36$.
A more detailed study of the general phase diagram for the
disordered and ordered phases as a
function of pinning strength versus temperature appears
in Ref.~\cite{Reichhardt22a}. 
The results in Fig.~\ref{fig:1} and
Fig.~\ref{fig:2} indicate that the system exhibits three distinct regimes.
These are an 
ordered or crystal regime containing sixfold-coordinated charges,
which occurs at low temperatures or low pinning strengths;
a disordered or plastic regime
where the system has low mobility and is strongly
affected by the pinning;
and a high temperature fluid phase where the effectiveness
of the pinning is reduced and the system
is in a strongly fluctuating state. 
In terms of transport, in the presence of pinning
the ordered state exhibits elastic depinning in which
the charges maintain their same neighbors. The glass  
state undergoes plastic depinning,
and the fluid state does not have a pinned
phase but can still have a regime in which
the charges are trapped for a time
before thermally hopping out of the pinning sites.

\section{Transport and Noise in the Plastic Regime} 
We next examine the noise
and transport in the three regimes identified above. 
We
consider samples with
$F_{p}= 0.5$,
a pinning strength at which the charges are disordered
for $T/T_{m} = 0$,
so the system
is in a strongly disordered glass phase.
The $T/T_{m} = 0$ plastic depinning
that occurs in this regime
was studied in detail in \cite{Reichhardt22a}, 
where a pinned phase, a filamentary flow phase, 
a disordered flow phase, and a dynamically ordered moving smectic phase
appear in sequence as a function of increasing drive.

\begin{figure}
\includegraphics[width=\columnwidth]{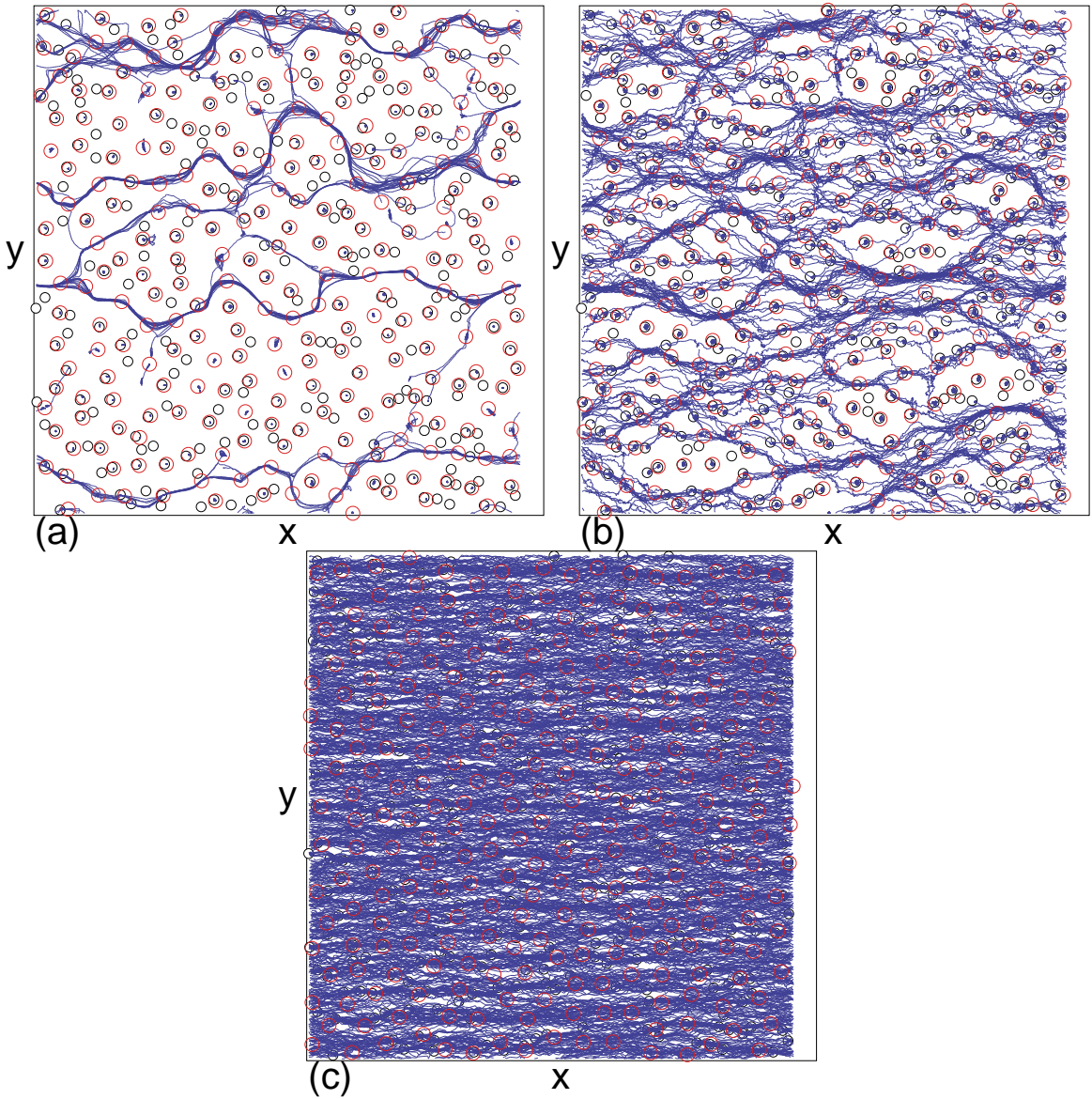}
\caption{
Charge locations (red circles), trajectories (blue lines),
and pinning site locations (black circles)
for the system in Fig.~\ref{fig:2} at 
$F_{D} = 0.15$ and $F_{p}= 0.5$. 
(a) Filamentary flow at $T/T_{m} = 0.15$.
(b) Disordered flow with channels at $T/T_{m} = 0.606$. 
(c) $T/T_{m} = 1.03$.}
\label{fig:3}
\end{figure}

In Fig.~\ref{fig:3}(a) we show 
a snapshot of the charge locations, pinning site locations,
and trajectories in the plastic flow regime
for $F_{D}= 0.15$ at $T/T_{m} = 0.15$, where a portion of the charges 
are moving in a series of well defined channels,
with occasional jumps 
between the channels when certain channels open or close again.
In general, for
strong pinning, at low temperature the system exhibits 
channel flow just 
above depinning,
similar to that studied in other systems at zero temperature. 
Figure~\ref{fig:3}(b) shows the same system
at $T/T_{m} = 0.606$ where there is a combination
of channel flow and random thermal hopping, indicating that as the
temperature increases, there is a
transition from one-dimensional (1D) channels to
two-dimensional (2D) flow. 
In Fig.~\ref{fig:3}(c), at $T/T_{m}= 1.03$ 
the motion is 2D and fluidlike.
For higher temperatures, the images look similar
to what is shown in Fig.~\ref{fig:3}(c). 

 \begin{figure}
\includegraphics[width=\columnwidth]{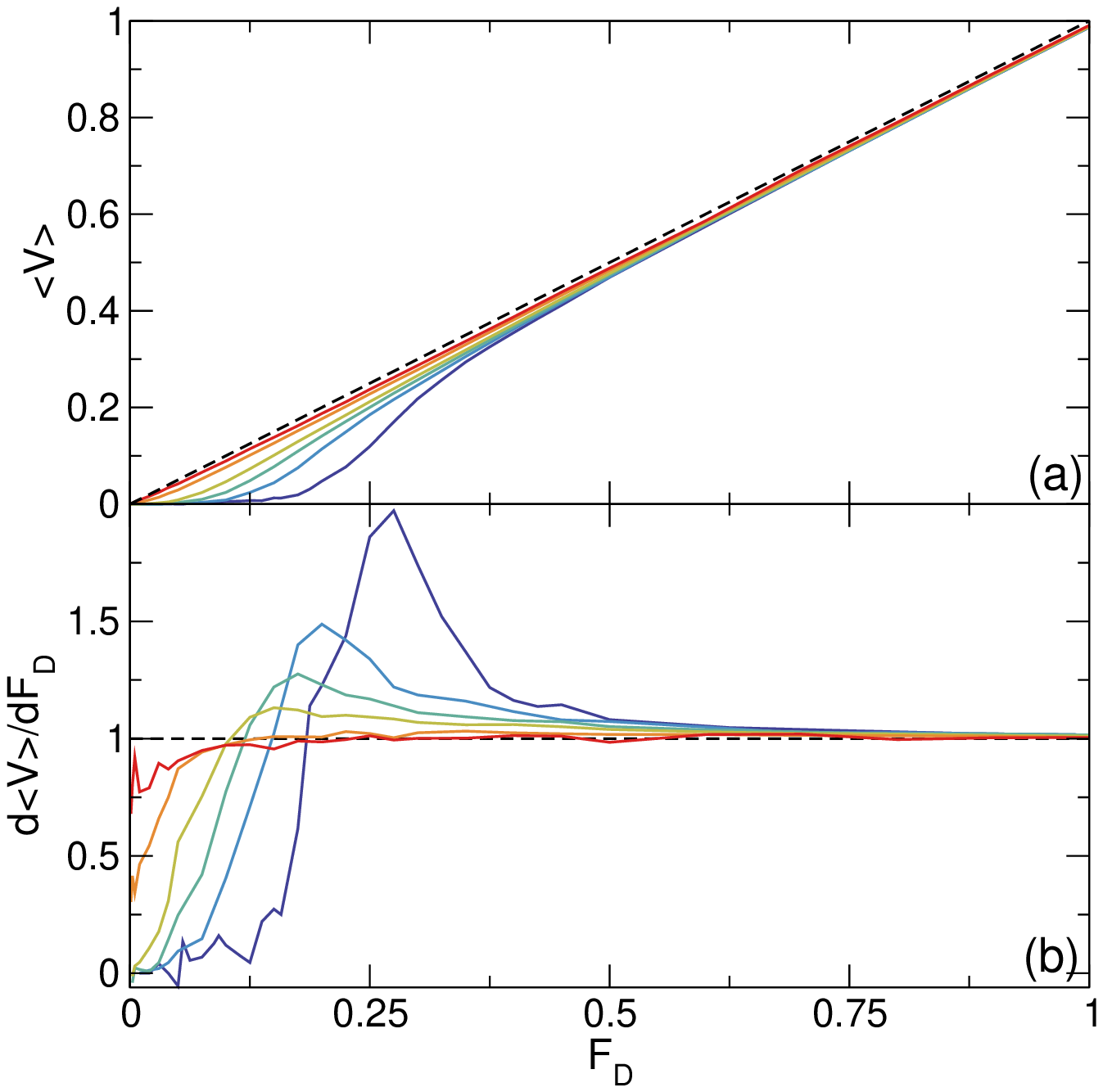}
\caption{
(a) Velocity $\langle V\rangle$ vs drive $F_{D}$ 
for the system in Fig.~\ref{fig:3} with
$F_p=0.5$ at
$T/T_{m} = 0.0$ (purple), 
$0.606$ (blue), $0.91$ (dark green), $1.21$ (light green), $1.81$
(orange), and $2.42$ (red).
(b) The corresponding $d\langle V\rangle/dF_{D}$ vs $F_D$ curves.   
The dashed lines indicate the pin-free limit.
}
\label{fig:4}
\end{figure}

In Fig.~\ref{fig:4}(a) we
plot $\langle V\rangle$ versus $F_{D}$ for the system in
Fig.~\ref{fig:3} at
$T/T_{m} = 0$, $0.606$, $0.9$, $1.21$,
$1.81$, and $2.42$.
For the lower temperatures, there is
a well-defined
depinning threshold followed by a nonlinear regime,
while when $T/T_m > 1.0$, 
the threshold is replaced by a creep regime
and the nonlinear regime at higher drives persists.
At the highest drives, all of the curves approach the pin-free limit. 
In Fig.~\ref{fig:4}(b) we show the corresponding
$d\langle V\rangle/dF_{D}$ versus $F_D$ curves, where
for $T/T_{m} \geq 1.21$ there is a peak 
in $d\langle V\rangle/dF_{D}$ due to the S shape
of the velocity-force curves.
Similar peaks in the differential conductivity were observed for
driven superconducting vortices in the 
plastic flow regime \cite{Bhattacharya93,Olson98a,Hellerqvist96,Ryu96}.
For $T/T_{m} > 1.21$, the
peaks are lost and a creep regime appears.
The dashed line is the
differential conductivity for the pin free system,
and all of the curves approach this value at high drives.
We note that for $T/T_{m} = 0$, at low 
drives there are a number of jumps in the conduction as well as
a few regimes of negative differential conduction.
This arises due to a filamentary flow channel effect that is described
in more detail in \cite{Reichhardt22}.
For $T/T_{m} > 0.5$, the jumps 
associated with the filamentary flow phase are lost 
and a single large peak in $d\langle V\rangle/dF_{D}$
appears in the plastic flow regime where there is a combination of
moving and pinned charges.

\begin{figure}
\includegraphics[width=\columnwidth]{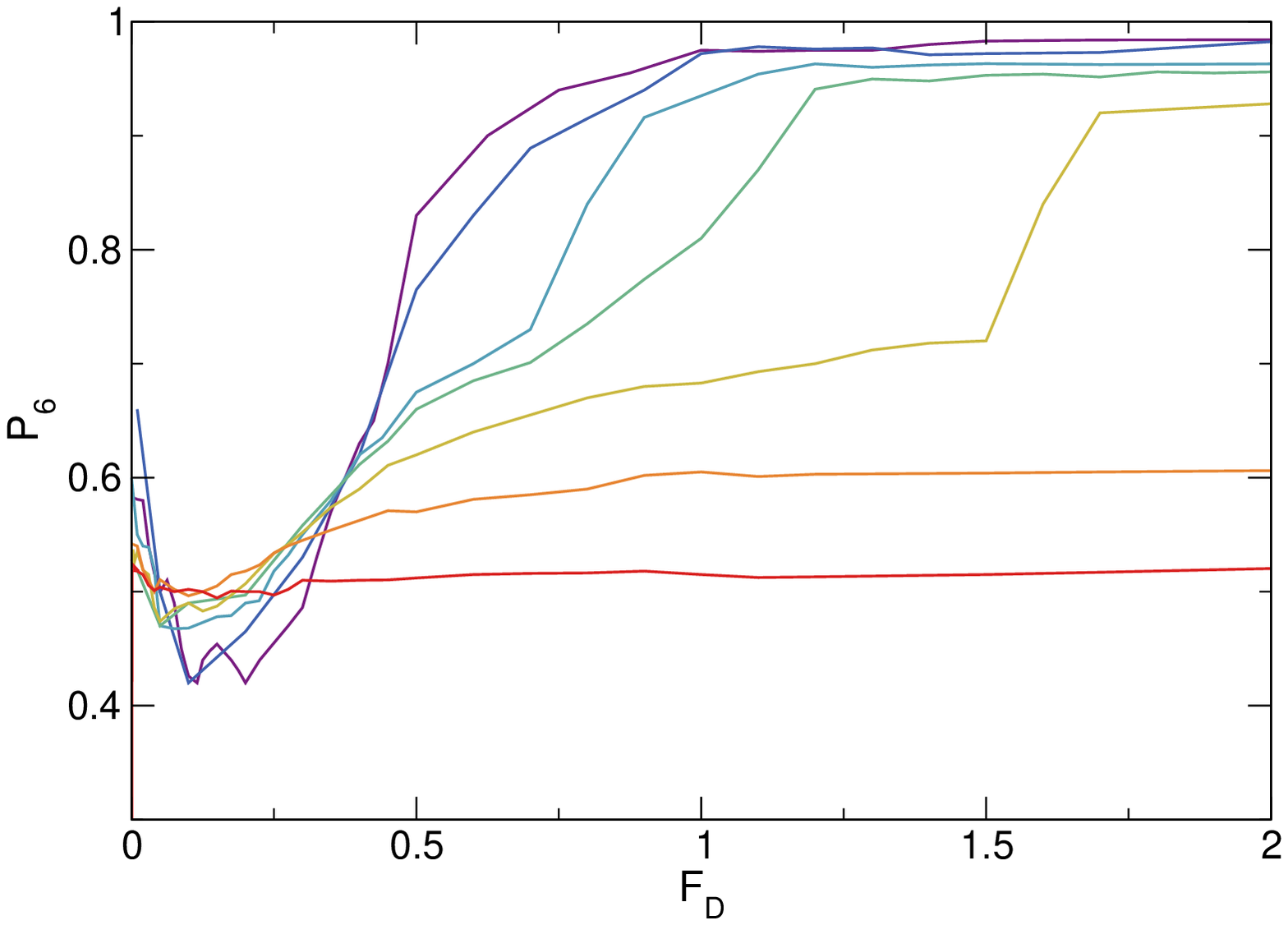}
\caption{
$P_{6}$ vs $F_{D}$ for the system in Fig.~\ref{fig:4}
with 
$F_p=0.5$ at
$T/T_{m} = 0$ (purple), $0.303$ (dark blue), $0.606$ (light blue),
$0.757$ (green), $0.91$ (yellow), $1.21$ (orange),
and $2.42$ (red).  The system reaches
an ordered state at high drives for $T/T_{m} < 1.0$.
}
\label{fig:5}
\end{figure}

In Fig.~\ref{fig:5} we plot $P_{6}$ versus $F_{D}$ for the system in
Fig.~\ref{fig:4} for 
$T/T_{m} = 0$, $0.303$, $0.606$, $0.757$,
$0.91$,
$1.21$, and $2.42$.
For $T/T_{m} < 1.0$ there is
an initial dip in $P_{6}$ at the onset of plastic flow, and at high
drives where $d\langle V\rangle/dF_{D}$
starts to flatten, $P_{6}$ approaches
values of $0.9$ or higher
as the system forms a moving smectic phase.
In the moving smectic state,
the charges move in well defined channels and a small number of
dislocations are present that have their Burgers vectors
aligned with the driving direction
\cite{Reichhardt17,Reichhardt01,Reichhardt22}. 
As $T/T_m$ increases further, the drive at which the smectic
state emerges shifts
to higher values of $F_{D}$,
and for $T/T_{m} > 1.0$, the system no longer
forms a moving smectic but instead becomes a moving fluid. 

 \begin{figure}
\includegraphics[width=\columnwidth]{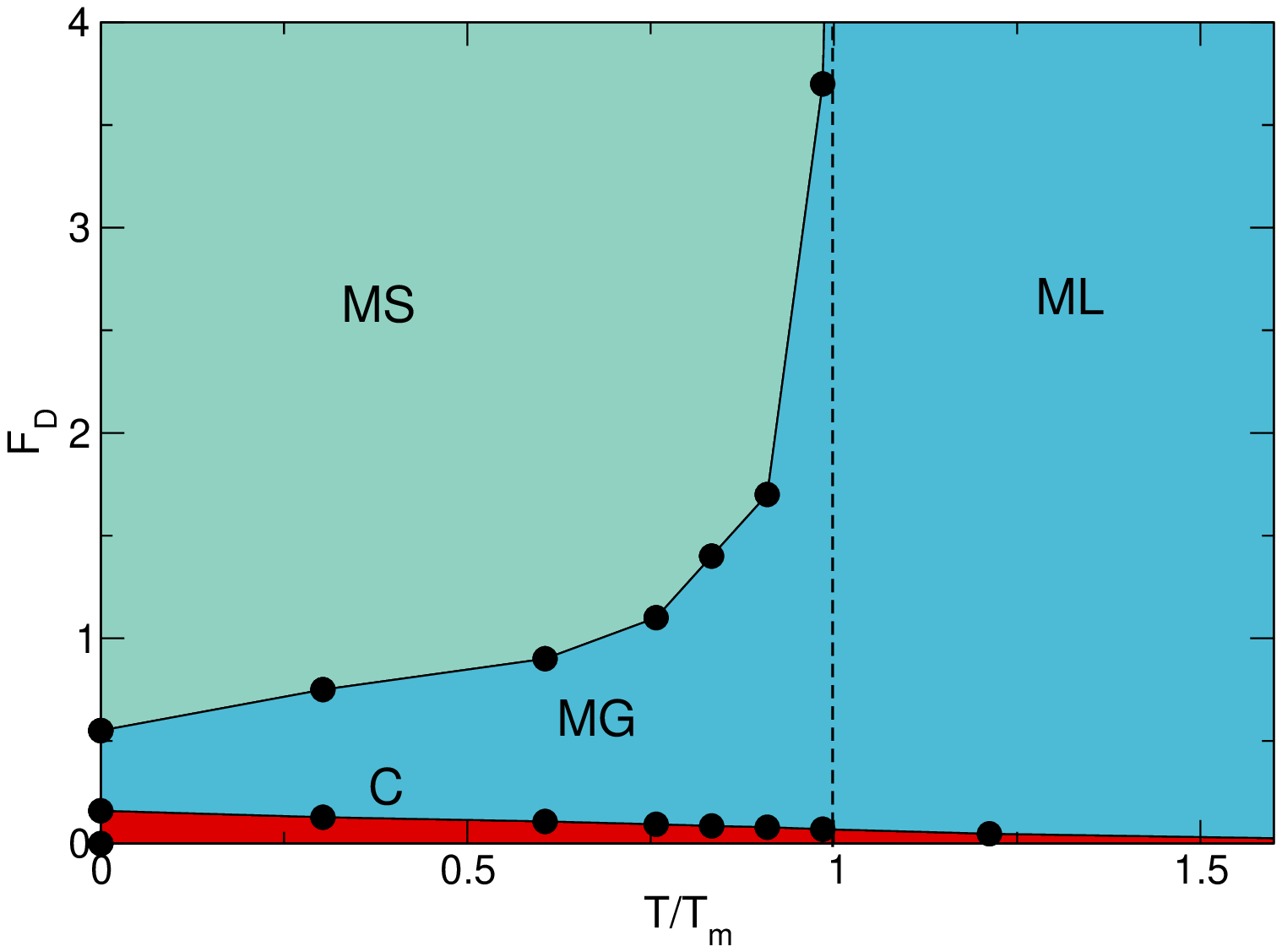}
\caption{
Dynamic phase diagram as a function of $F_{D}$ vs $T/T_{m}$
for the system in Figs.~\ref{fig:4} and \ref{fig:5}
with
$F_p=0.5$.
There is a pinned or creep phase C (red),
a disordered moving phase (blue) that is a moving glass, MG, at lower
temperatures and a moving liquid, ML, at higher
temperatures, and a moving smectic MS (green).
}
\label{fig:6}
\end{figure}

From the features in the transport curves and $P_{6}$ plotted
in Figs.~\ref{fig:4} and \ref{fig:5},
we construct a dynamic phase diagram of the evolution of the
different phases as a function of
$F_{D}$ versus $T/T_{m}$ in Fig.~\ref{fig:6}.
At low drives we find a pinned or creep regime
denoted C, where
$d\langle V\rangle/dF_{D} < 0.5$.
The dynamically ordered moving smectic phase MS
appears when $P_{6} > 0.9$. The disordered regime is where the
system is structurally disordered but moving,
and it can be either a moving glass MG
for $T/T_{m} < 1.0$, or a moving liquid ML for $T/T_{m} > 1.0$.
The overall features of the
phase diagram are similar 
to those observed in driven superconducting vortex systems
with quenched disorder,
as first proposed by Koshelev and Vinokur \cite{Koshelev94},
where the transition between the MG and MS
states shifts
to higher drives as $T/T_{m}$ is approached. 
In Ref.~\cite{Koshelev94},
the transition line from the disordered to moving ordered
phase was argued 
to be proportional to $A/(T_{m} - T)$, where $A$ is some prefactor and
the moving ordered phase can be described in terms of having an
effective temperature that is decreasing toward zero.
This picture assumes the formation of a moving crystal at high drive,
and is somewhat modified
in our system since the moving state we observe
is a smectic in which the
dynamic fluctuations are anisotropic \cite{Balents98}.
We find that a better fit to the transition line in our
case is $(T_{m} -T)^{-0.7}$,
which is likely due to the anisotropic nature of the moving smectic. 

Now that we have established the
dynamic phase diagram as a function of drive versus temperature,
we can ask how the
velocity fluctuation power spectra
change as a function of $F_{D}$ and $T$. 
The power spectrum
as a function of $\omega=2\pi f$ can be calculated using
the time series $v(t)$ of the velocity fluctuations,
\begin{equation}
S(\omega) = \left|\int v(t)e^{-i\omega t}\right|^2 
\end{equation}
At $T = 0$ the noise has a $1/f^{\alpha}$ 
signature with $\alpha \approx 0.8$, in agreement with
recent experiments \cite{Brussarski18}.
The noise power is reduced at high drives and 
shows a crossover to a narrow band signature when the system forms
a moving smectic phase;
however, the experiments do not
detect a narrow band noise signature
at higher drives,
suggesting that thermal effects could be coming
into play.  

\begin{figure}
\includegraphics[width=\columnwidth]{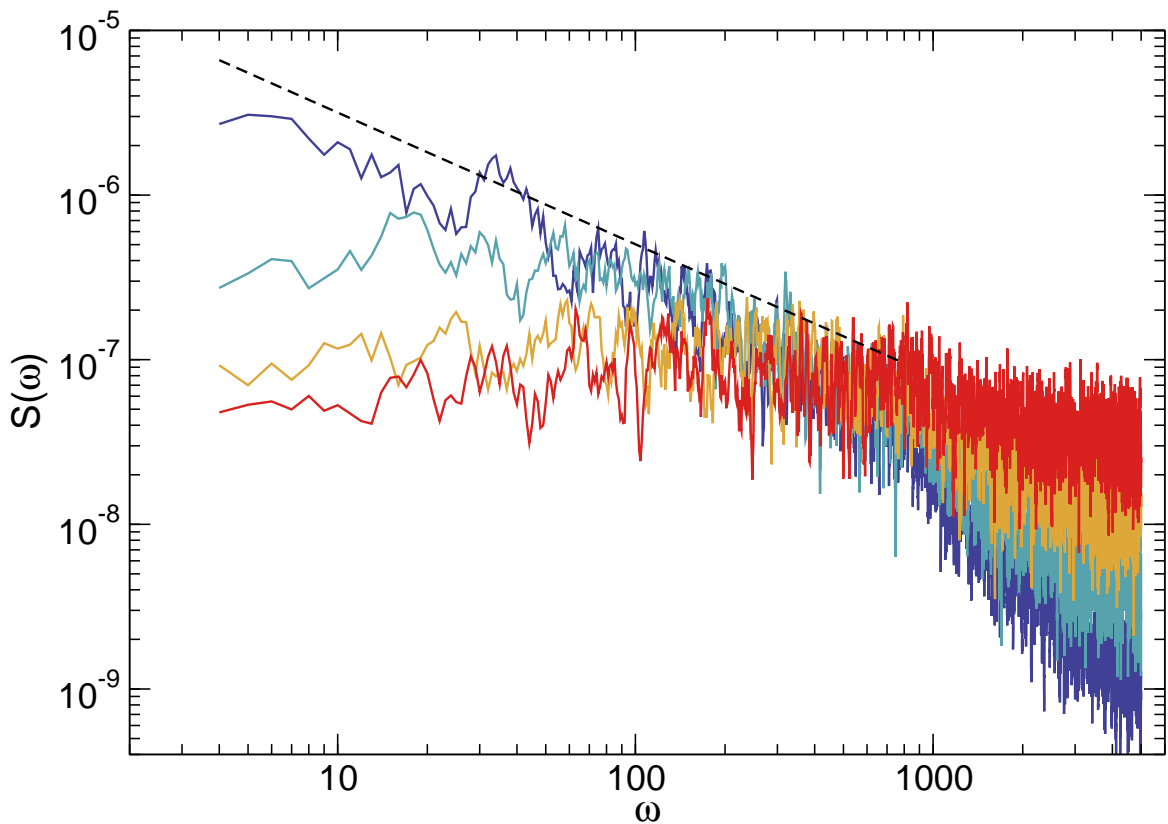}
\caption{
Power spectra $S(\omega)$ vs $\omega$ for the system in Fig.~\ref{fig:6}
with
$F_{D} = 0.15$ for
$T/T_{m} = 0$ (dark blue), $0.303$ (light blue), $0.606$ (yellow),
and $1.03$ (red). The spectral signature changes from
$1/f$ to white at low frequencies as the temperature increases,
while the amount of noise power at higher frequencies increases
with increasing $T$. 
}
\label{fig:7}
\end{figure}

In Fig.~\ref{fig:7} we plot power spectra of the
velocity time series for the system in
Fig.~\ref{fig:6} at a drive of $F_{D} = 0.15$ for 
$T/T_{m} = 0$, $0.303$, $0.606$, and $1.03$.
For $T/T_{m} = 0$, the low frequency noise
has a $1/f^{\alpha}$ form,
where the dashed line is a fit with $\alpha = -0.8$,
while at higher frequencies the noise tail has
$\alpha = -2.0$.
At $T/T_{m} = 0.5$, the lower frequency noise
power is reduced
and $\alpha$ drops closer to $\alpha=0$, the
value expected for white noise; however, the high
frequency noise still has a
$1/f^2$ form.
For higher $T/T_{m}$, the 
low frequency noise power is
further reduced while the higher frequency noise
power is enhanced, and the spectrum becomes much whiter overall.

\begin{figure}
\includegraphics[width=\columnwidth]{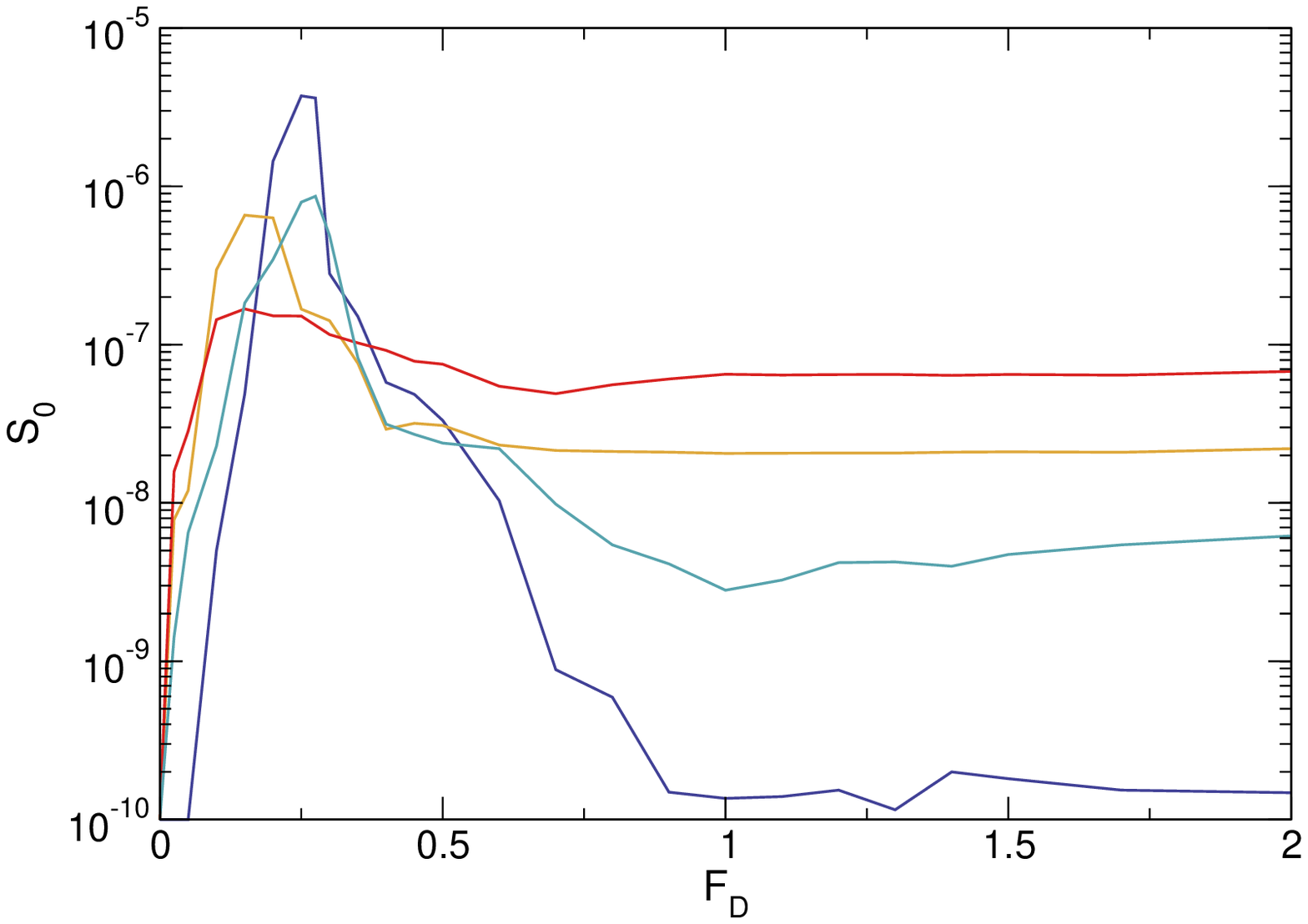}
\caption{
The noise power $S_{0}$ at fixed $\omega=20$
vs $F_{D}$
for the system in Fig.~\ref{fig:7}
at $F_{D} = 0.15$ for
$T/T_{m} = 0$ (dark blue), $0.303$ (light blue),
$0.606$ (yellow), and $1.03$ (red).
}
\label{fig:8}
\end{figure}

To better characterize the system,
we measure the noise power $S_0$, which is
the value of the spectral power integrated in a small window
around a specific frequency $\omega=20$.
In Fig.~\ref{fig:8}
we show $S_{0}$ versus $F_{D}$ for $T/T_{m} = 0$, $0.303$,
$0.606$, and $1.03$ on a log-linear plot.
For $T = 0$ there is a large peak 
in $S_0$ over the range $0.01 < F_{D} < 0.5$,
which corresponds to the appearance of $1/f^{0.85}$ noise.
The noise is white for $0.5 < F_{D} < 0.9$,
and for $F_{D} > 0.9$ a narrow band noise signal
appears.
For $T/T_{m} = 0.303$ and $0.606$, there is still 
a peak in the noise near $F_{D} = 0.2$, but as the temperature
increases, the peak power diminishes and the peak
location shifts to lower drives.
This is correlated with a whitening of the low frequency noise,
as shown in Fig.~\ref{fig:7}.
For $T/T_{m} = 1.03$, the sharp noise power peak is lost.
At large $F_D$, we find that the noise power increases with
increasing temperature due to the transition
from flow through narrow 1D channels in the
smectic state to a 2D Brownian like motion
in the liquid state.
The overall behavior of the noise power that we find
is in agreement
with experimental observations
\cite{Brussarski18}, where there is a large peak in the noise power 
near the depinning threshold at low temperatures, 
while for higher temperatures
the noise power peak is reduced and shifts to lower drives
before disappearing
at sufficiently high temperature.
Another feature
that is also observed in the experiments
is that the noise power increases with temperature
at large drives.

\begin{figure}
\includegraphics[width=\columnwidth]{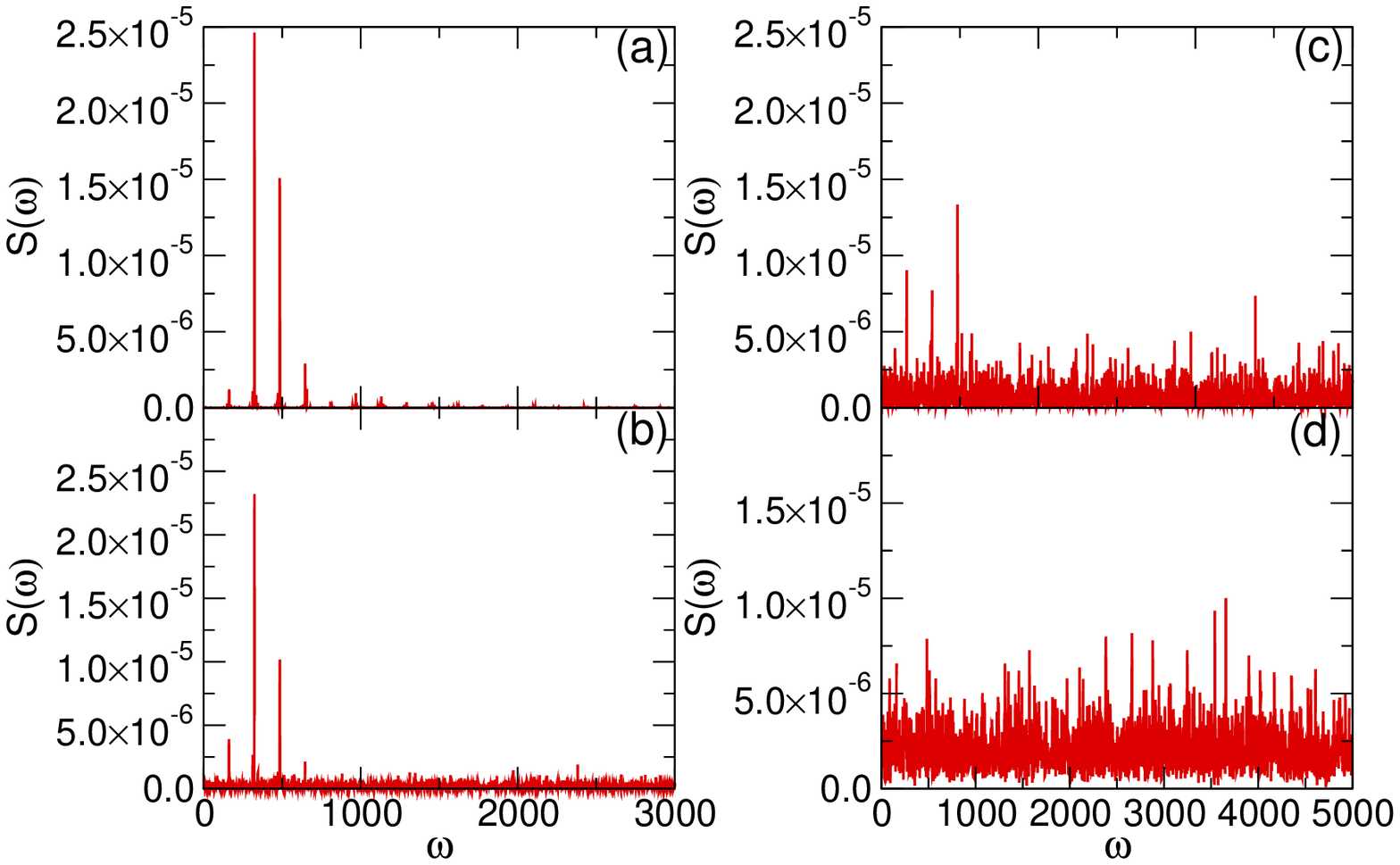}
\caption{
$S(\omega)$ vs $\omega$ for the system in
Fig.~\ref{fig:6}
at 
$F_{D} = 1.5$ where the system is in the moving smectic phase. 
$T/T_m=$ (a) 0, (b) $0.303$, (c) $0.606$, and (d) $1.03$.  
}
\label{fig:9}
\end{figure}

\begin{figure}
\includegraphics[width=\columnwidth]{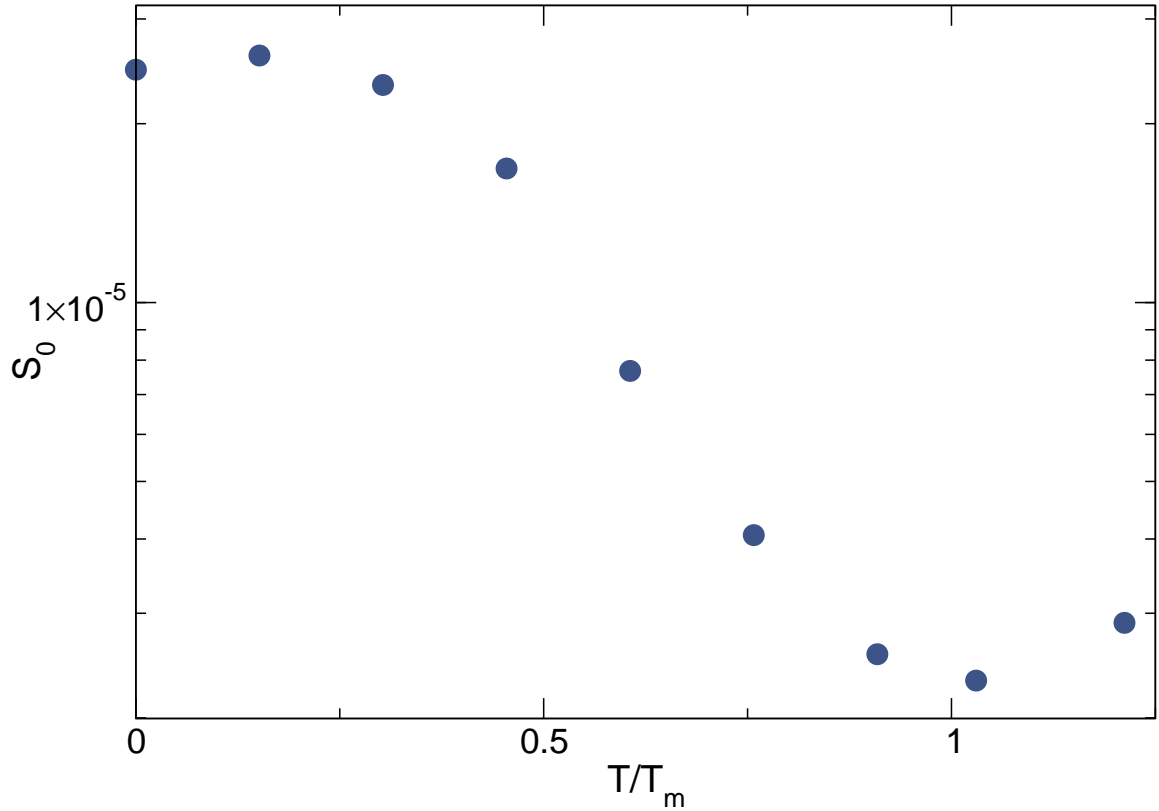}
\caption{
The value of the noise power $S_0$ at $\omega=323$, the frequency of the
largest narrow band noise peak in Fig.~\ref{fig:9}, vs $T/T_m$
for the system in Fig.~\ref{fig:8} with
$F_D=0.15$.
Here the narrow band noise
peaks are lost near $T/T_{m} = 0.75$. 
}
\label{fig:10}
\end{figure}

\begin{figure}
\includegraphics[width=\columnwidth]{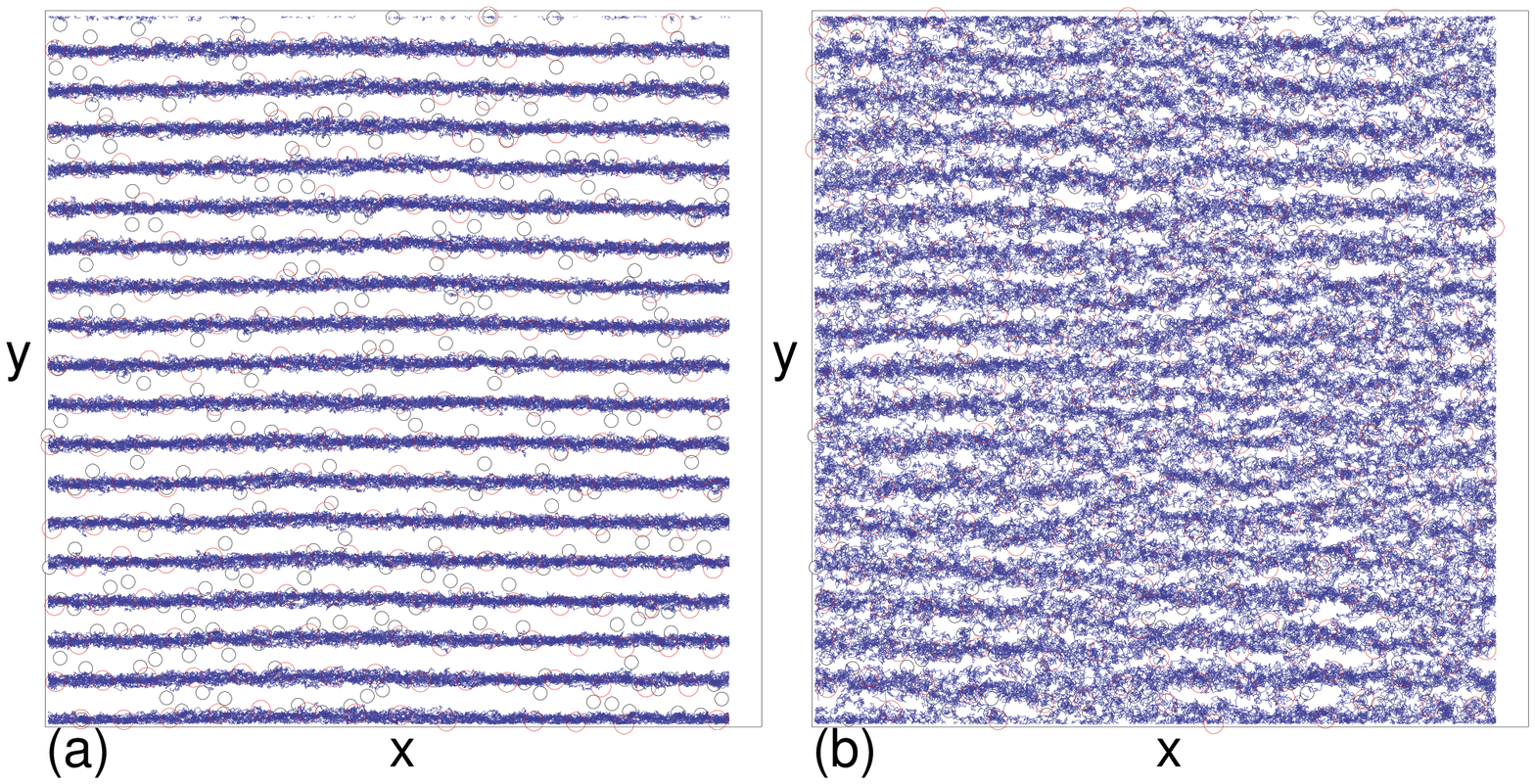}
\caption{
Charge locations (red circles), trajectories (blue lines), and pinning
site locations (black circles) 
for the system in Figs.~\ref{fig:9} and \ref{fig:10}
at $F_{D} = 1.5$ for
$T/T_{m} =$ (a) 0.606 and (b) $1.03.$  
}
\label{fig:11}
\end{figure}

We next consider thermal effects in the high drive limit where
the system forms a moving smectic at $T=0$.
In Fig.~\ref{fig:9}(a) we show $S(\omega)$ vs
$\omega$ for the system from Fig.~\ref{fig:6} at $F_{D} = 1.5$ and
$T/T_{m}= 0$, where there are a series of peaks
associated with a narrow band noise signature.
At $T/T_m=0.303$ in Fig.~\ref{fig:9}(b),
there are still strong peaks associated with the
narrow band noise but
the higher harmonic peaks are strongly reduced in power.
In Fig.~\ref{fig:9}(c) at $T/T_{m} = 0.606$, the level of background
noise has increased
and the narrow band peaks are diminished in size,
while at $T/T_{m}= 1.03$ in Fig.~\ref{fig:9}(d),
the moving smectic phase is lost and the
narrow band peaks disappear into the background noise.
To better characterize the change in the
narrow band noise signature,
in Fig.~\ref{fig:10} we plot the noise power $S_0$ at $\omega=323$, which
is the location of the most pronounced narrow band noise peak
in Fig.~\ref{fig:9}(a). 
For $T/T_{m} < 0.5$ there is a strong narrow band noise signal; however,
at $T/T_{m} = 0.75$ the narrow band noise
level is close to the value of the 
background noise.
This suggests that thermal effects can strongly reduce the narrow band
noise signal even at temperatures well below $T/T_{m}= 1.0$, which could
explain why the narrow band noise signals are
difficult to see in experiment. 
To better understand the
origins of the changes in the noise signals,
in Fig.~\ref{fig:11}(a) we plot the
trajectories of the charges at $T/T_{m}= 0.606$
where narrow band noise is present.
The system is still in a moving smectic state but the channels 
have been broadened by the thermal fluctuations,
and there are several regions in which the
channel structures are starting to break down.
Figure~\ref{fig:11}(b) shows the trajectories for 
$T/T_{m} = 1.07$, where the 1D channel structure is lost,
there is a significant amount of transverse diffusion, and the narrow
band noise peaks disappear.
This result indicates that the narrow band noise
occurs only when the motion of the charges is mostly 1D in character.

\begin{figure}
\includegraphics[width=\columnwidth]{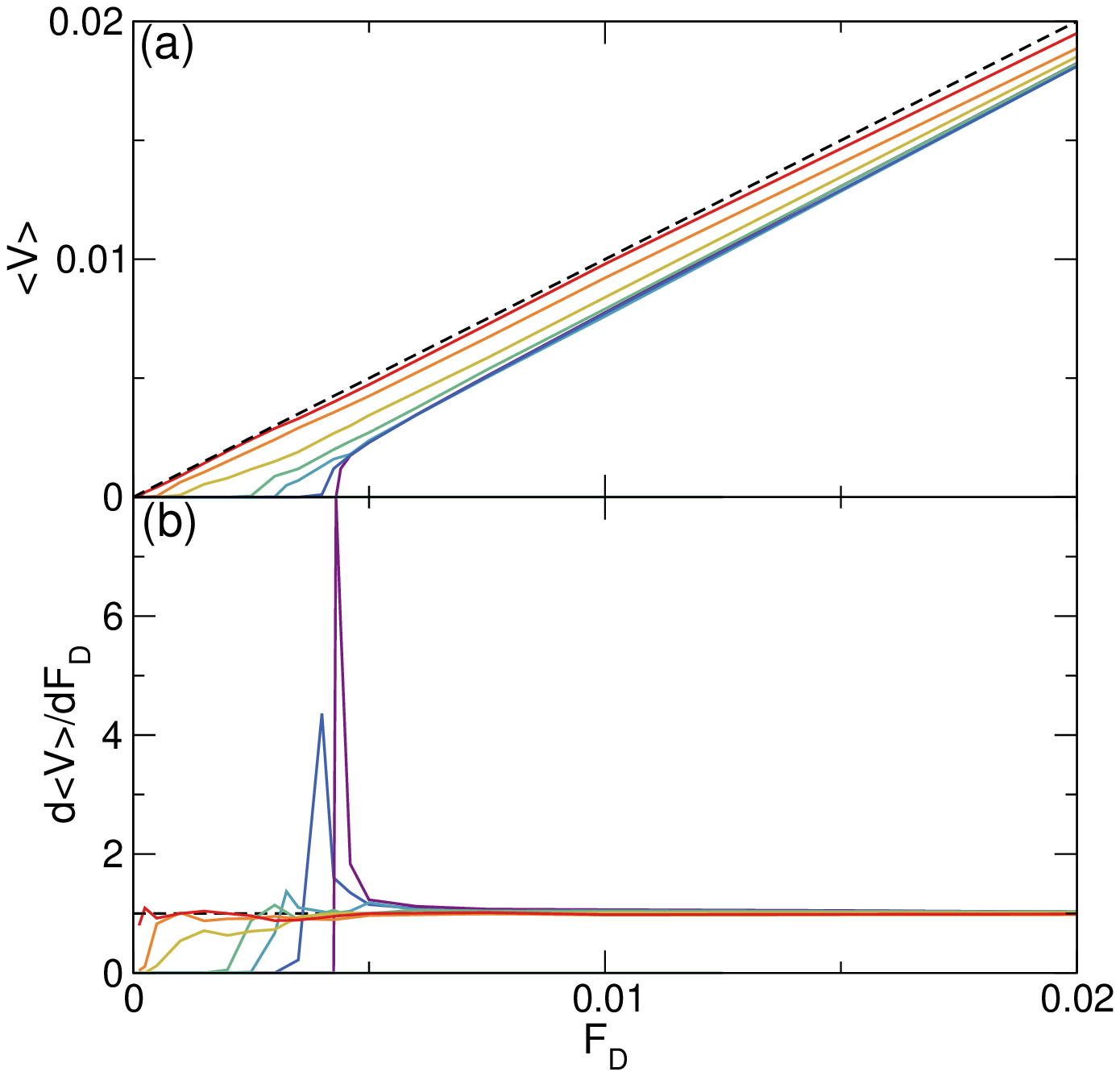}
\caption{(a) $\langle V\rangle$ vs $F_{D}$
for a system that exhibits elastic depinning at $T/T_{m} = 0$, where
$F_{p} = 0.05$.
The different curves are for temperatures of
$T/T_m = 0$, 0.0378, 0.0756, 0.17, 0.303, 0.606, and $1.03$,
from right to left.
The dashed line is the expected curve in the pin free limit.
(b) The corresponding $d\langle V\rangle/dF_{D}$ vs $F_D$ curves. 	
}
\label{fig:12}
\end{figure} 

\section{Thermal Depinning Noise in the Elastic Regime}
We next consider the thermal depinning and noise
in the elastic regime where the
charges maintain their same neighbors.
From Fig.~\ref{fig:2} we select a value of 
$F_{p} = 0.05$, well below the $T/T_m = 0$ disordering
threshold of $F_{p} = 0.075$.
In Fig.~\ref{fig:12}(a) we show
$\langle V\rangle$ versus $F_{D}$ at $F_{p} = 0.05$
for $T/T_m = 0$, 0.0378, 0.0756, 0.17, 0.303, 0.606, and $1.03$, and
we plot the corresponding
$d\langle V\rangle/dF_D$ curves in Fig.~\ref{fig:12}(b).
As $T/T_{m}$ increases, the depinning threshold
shifts to lower $F_D$,
and in Fig.~\ref{fig:12}(b), the peak in $d\langle V\rangle/dF_D$ that appears
for $T = 0$ 
is lost for $T/T_m >0.303$.
We note that the system remains in an
ordered state up to $T/T_{m} = 1.0$ for all values
of $F_{D}$. The $d\langle V\rangle/dF_{D}$ curves also show a
multiple peak feature at high temperatures, with one peak at
the finite temperature threshold and a second peak
near the $T = 0$ depinning threshold. In between these two peaks,
the flow is creep-like in nature.

\begin{figure}
\includegraphics[width=\columnwidth]{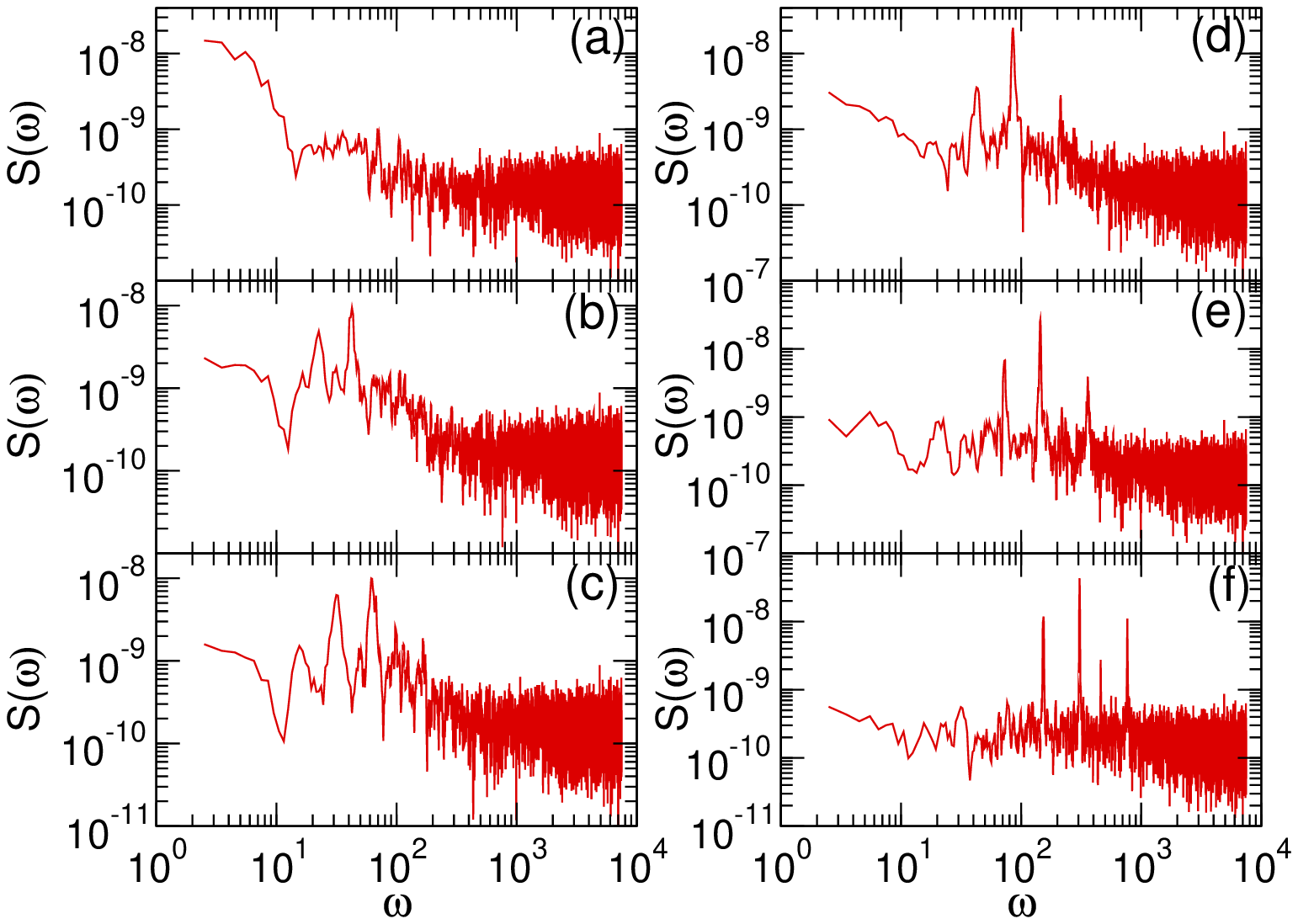}
\caption{
$S(\omega)$ vs $\omega$ for the system in Fig.~\ref{fig:12}
with
$F_p=0.05$
at $T/T_{m} =0.1515$ 
for $F_{D} =$ (a) 0.025, (b) 0.03, (c) 0.04, (d) 0.046, (e) 0.06,
and (f) $0.01$.
}
\label{fig:13}
\end{figure}

In Fig.~\ref{fig:13} we plot $S(\omega)$ versus $\omega$
for the system in Fig.~\ref{fig:12} 
at $T/T_{m} = 0.1515$ for different values 
of $F_{D} = 0.025$, 0.03, 0.04, $0.046$, $0.06$, and $0.01$. 
At $F_{D}=0.025$, the motion occurs mostly in the form 
of avalanches,
and no clear narrow band signatures are present
but the low frequency noise has high power.
For $F_{D} = 0.03$, the system starts to develop
a narrow band noise signature that sharpens with increasing drive,
and for $F_{D} \geq 0.06$, which is above the zero 
temperature depinning threshold,
the low frequency noise is strongly suppressed and the
narrow band noise peaks become much sharper.
This result shows that in the elastic
flow regime,
the narrow band noise signal is more robust
than in the plastic phase,
and it appears once the system has depinned. 

\begin{figure}
\includegraphics[width=\columnwidth]{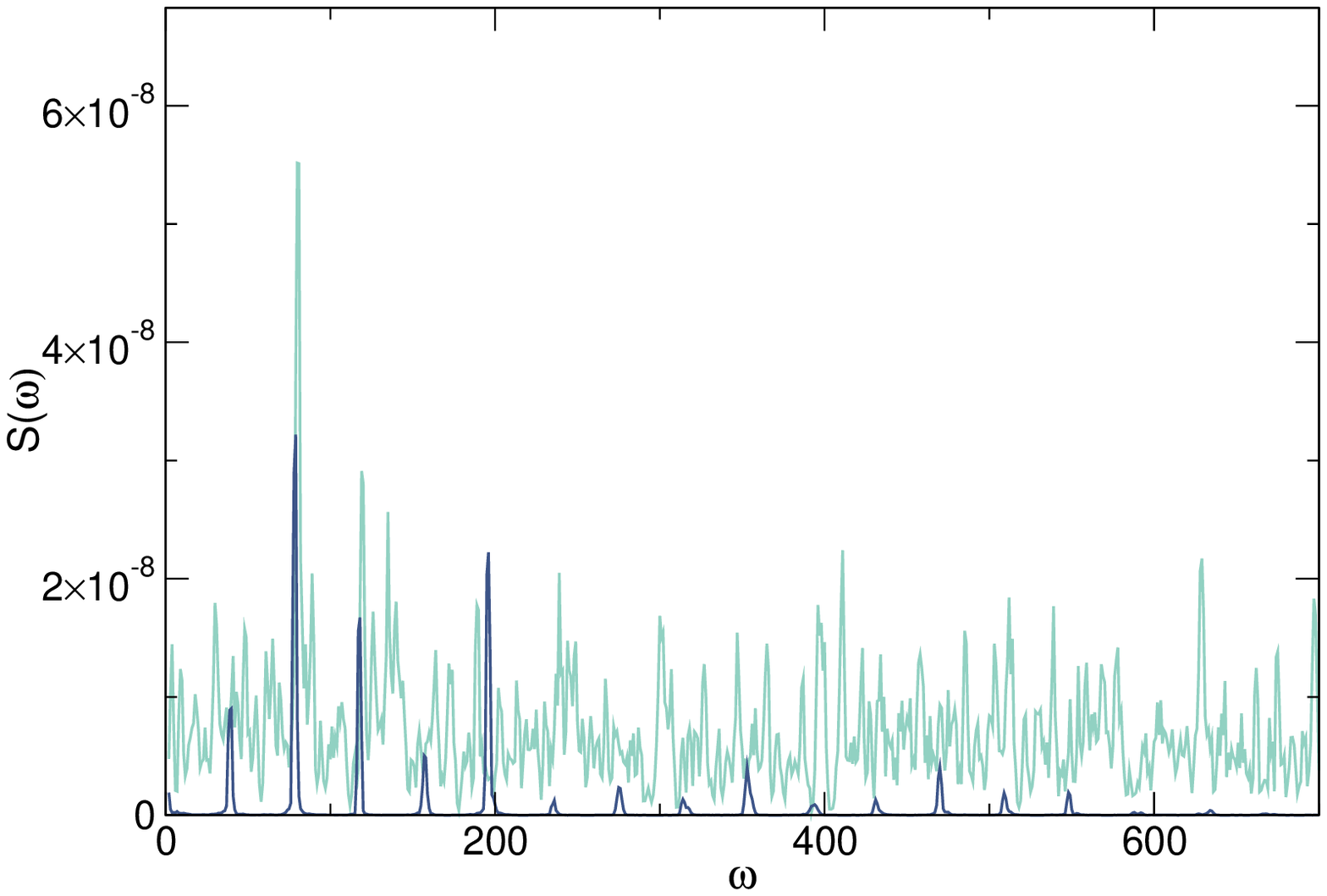}
\caption{
The power spectra $S(\omega)$ vs $\omega$ for the system in
Fig.~\ref{fig:13} with
$F_p=0.05$ at 
$F_{D} = 0.02$ in the moving phase for $T/T_{m} = 0$ (blue) and
$T/T_{m} = 0.303$ (green).
At $T/T_{m} = 0.303$, although the overall background noise
power is higher, there is an enhancement of the narrow band noise
signal. 
}
\label{fig:14}
\end{figure}

In Fig.~\ref{fig:14} we plot $S(\omega)$ vs $\omega$ for the system in 
Fig.~\ref{fig:13} at $T/T_{m} = 0$ and $T/T_m = 0.303$
at a drive of $F_{D} = 0.02$. 
At $T/T_{m} = 0$, there is a strong narrow band noise feature.
Interestingly, at $T/T_m = 0.303$, although the level of background noise
has increased, the primary narrow band noise peak is enhanced in power.
The increase in the narrow band peak occurs when
thermal effects weaken the effectiveness of the pinning
and allow the charges to become better ordered. This effect
is diminished in the case of strong pinning.

\begin{figure}
\includegraphics[width=\columnwidth]{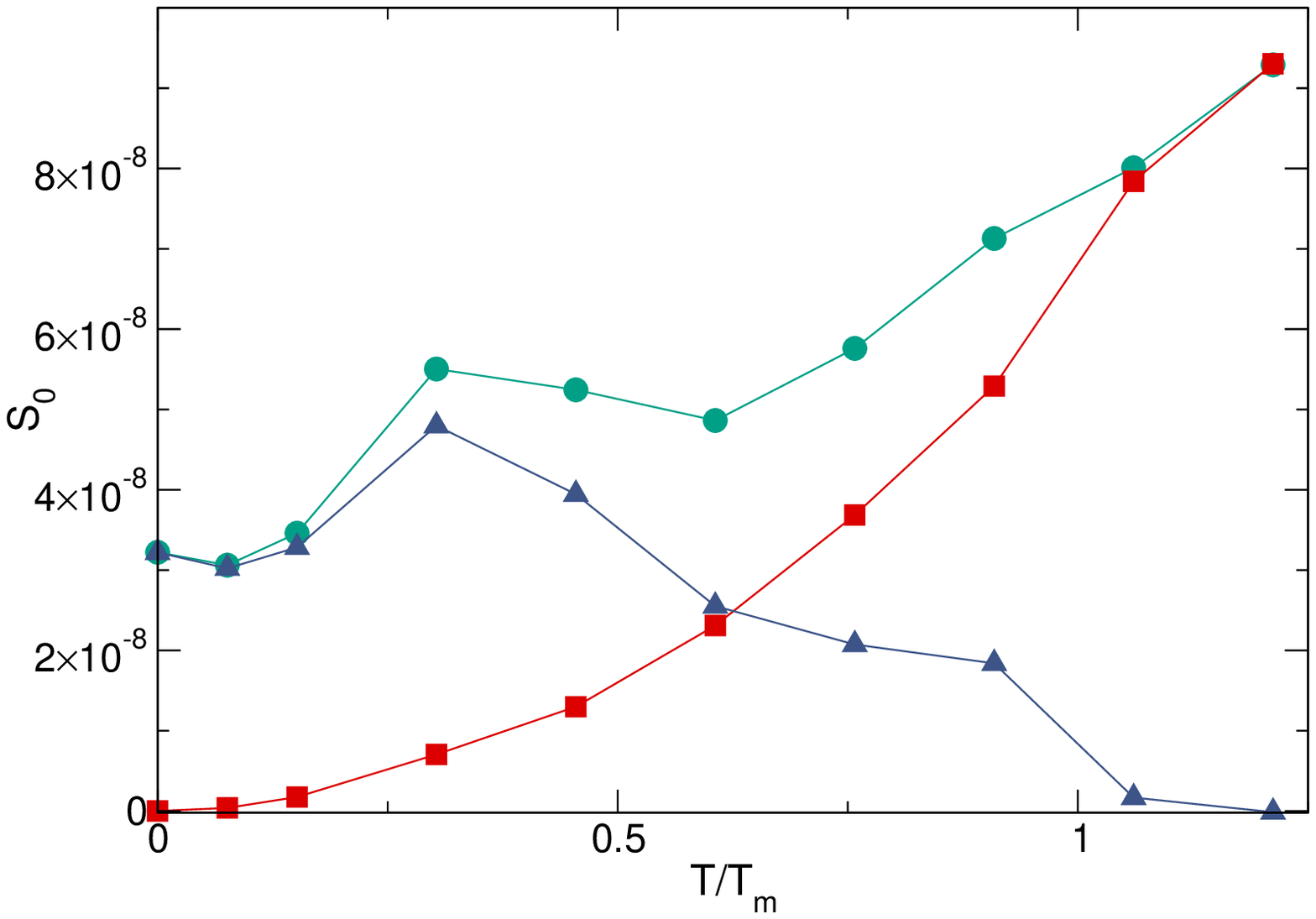}
\caption{
The noise power $S_0$ vs $T/T_{m}$ for the system in Figs.~\ref{fig:12}
and \ref{fig:13} with
$F_p=0.05$ at $F_D=0.02$
for the narrow band frequency of
$\omega=80$ (green circles), the background
noise at $\omega=300$
(red squares), and the difference (blue triangles).  
}
\label{fig:15}
\end{figure}

To better characterize the narrow band noise
behavior for the system in Figs.~\ref{fig:12} and \ref{fig:13},
in Fig.~\ref{fig:15} we plot the noise power
$S_0$ versus $T/T_m$ for $F_D=0.02$ where the system is always in
a moving state
at
the narrow band peak of $\omega = 80$
and the background noise signal at $\omega=300$,
along with the difference between these two noise powers.
Unlike the case for strong pinning,
the power of the narrow band
noise signal generally increases with increasing $T/T_{m}$; 
however, the background noise power also increases,
and the amount of power in the two signals becomes equal
near $T/T_{m} = 1.0$.
The narrow band noise peak has the greatest amount of additional power
compared to the background noise
near 
$T/T_{m}=0.3$.
This is again due to thermal effects washing out
any additional 	avalanche-like motion
and permitting the charge lattice to become better ordered.

\section{Discussion}

Narrow band noise has been observed experimentally
in superconducting vortex \cite{Okuma07,Okuma08},
magnetic skyrmion \cite{Sato19}, and
charge density wave \cite{Gruner88} systems, but has not been seen
for Wigner crystals.
There have been reports of periodic noise in charge ordering
systems such as stripe or bubble forming states
\cite{Sun22,Cooper03}; however, this noise generally appears at
low frequencies
and is probably not associated with the lattice-scale narrow band noise,
but instead arises due to the motion of some other
periodically moving
macroscopic scale
structure.
In the experiments of 
Brussarski {\it et al.} \cite{Brussarski18},
the peak noise power
decreased with increasing temperature, similar to what we observe, but
no narrow band noise signal was observed.
This could be the result of several possible factors.
If the drive applied to the system
is not uniform,
there could still be 
strong plastic flow at low drives; however,
at high drives the system may not 
form a uniformly ordered moving state
but could instead break into several
locally ordered regions
that are moving at different speeds.
Related to this, if the quenched disorder
has a wide range of strength so that some of
the charges are moving while
a small number remain pinned,
a disordered flow regime would emerge in which
narrow band noise is absent.
A narrow band noise signal could also be masked
by strong background noise.
In this case, the signal could be 
boosted by applying an additional ac drive. If the frequency of this
ac drive is swept, phase locking or Shapiro steps would appear when
the frequency comes into resonance with the narrow band signal
\cite{Okuma07}. 
Another possible issue is that the narrow band frequency could be too
high to detect with the available experimental setup;
however, for a system in the elastic depinning
limit, fairly low frequency periodic signals could 
be generated in the creep regime.
The lack of experimentally observed narrow band noise
may suggest that elastic depinning of the Wigner crystal
is not occurring
and that the systems are generally in
the disordered or plastic flow regimes where the only available
narrow band noise signals are of the moving smectic type.
In principle, we think that the best place to look for
a narrow band noise signature is in a sample with relatively weak
pinning just above the depinning threshold. 
In our work, we focused on samples that were entirely
within the elastic or plastic regimes;
however, close to the transition between the elastic and plastic regimes,
the plastic flow noise may be enhanced.

\section{Summary} 

We have investigated the thermally induced depinning and noise fluctuations for
driven Wigner crystal systems with quenched disorder.
We identify an elastic regime in which the charges maintain the same
neighbors at depinning as well as a plastic regime in which
the system is broken up into moving and non-moving regions.
In the plastic depinning regime,
the velocity noise has a $1/f$ shape and there is a peak in the
noise power above the depinning threshold at lower temperatures,
while for large temperatures, the noise power peak is
reduced and the spectrum becomes white,
in agreement with experiments.
For high drives at low temperatures
in the plastic regime,
the system forms a moving smectic with a narrow band noise signal.
We find that this narrow band signal persists up
to $T/T_m = 0.75$, where $T_{m}$
is the temperature at which the charge lattice melts in the absence
of quenched disorder.
In the elastic regime, the system remains
ordered up to temperatures approaching $T/T_{m}= 1.0$,
although thermal effects reduce the depinning threshold.
In the elastic regime, $1/f$ noise appears
only in the creep regime where
there are avalanches or jumps of motion,
while in the sliding regime, pronounced narrow band noise appears
that reaches its lowest power
at the disorder-free melting temperature.
Our results show that measurements of the velocity noise spectra
and noise power
can be used in connection with transport curves
to distinguish different phases of driven Wigner crystals.

\ack
We gratefully acknowledge the support of the U.S. Department of
Energy through the LANL/LDRD program for this work.
This work was supported by the US Department of Energy through
the Los Alamos National Laboratory.  Los Alamos National Laboratory is
operated by Triad National Security, LLC, for the National Nuclear Security
Administration of the U. S. Department of Energy (Contract No. 892333218NCA000001).

\bibliographystyle{hunsrt.bst}
\bibliography{mybib}


\end{document}